\documentclass[twocolumn, prl,showpacs]{revtex4}

\usepackage{graphicx}
\usepackage{amssymb}
\begin{document}

\title{Stable fourfold configurations for small vacancy clusters
in silicon from ab initio calculations}

\author{D. V. Makhov}

\author{Laurent J. Lewis}

\affiliation{D\'epartement de physique
et Regroupement qu\'eb\'ecois sur les mat\'eriaux de pointe (RQMP)\\
Universit\'e de Montr\'eal,
Case Postale 6128, Succursale Centre-Ville,
Montr\'eal, Qu\'ebec H3C 3J7, Canada}

\begin{abstract}

Using density-functional-theory calculations, we have identified new stable configurations
for tri-, tetra-, and penta-vacancies in silicon.
These new configurations consist of combinations of a
ring-hexavacancy with three, two, or one interstitial atoms, respectively, such that all
atoms remain fourfold. As a result, their formation energies are lower by 0.6, 1.0, and
0.6 eV, respectively, than the ``part of a hexagonal ring''
configurations, believed up to now to be the lowest-energy
states.

\end{abstract}


\maketitle

Vacancies and their clusters are fundamental defects of silicon.
Usually they result from the irradiation of silicon with electrons \cite{Watkins1, el1, Motoko, el2},
neutrons \cite{n1,n2,n3}, protons \cite{pr2, Poirier}, or
ions \cite{ion1,ion2}, or from plastic deformations \cite{pl1, pl2}. However, vacancy clusters can also
be present in as-grown crystals \cite{asg}.
The presence of defects in crystalline semiconductors determines, to a large extent, their electrical
and optical properties,
making their study of great importance.

Calculations performed using density-functional-theory (DFT) molecular
dynamics \cite{Hast, Estr}, the Hartree-Fock method \cite{Hast, Estr},
and the DFT tight-binding method \cite{Staab}, among others, predict that the
ring hexavacancy should be significantly more stable than other types of vacancy clusters.
This can be explained using simple bond counting arguments: the crystal can reconstruct
almost perfectly around a hexavacancy, making all atoms remain fourfold.
For smaller clusters, the same calculations \cite{Hast, Estr, Staab} conclude that the most stable
configurations occur when atoms are removed sequentially from the hexagonal ring.

The ring hexavacancy is known to be a good trap for various impurities,
such as carbon, oxygen, and copper atoms \cite{Estr}. It is reasonable
to expect, therefore, that it may also be an efficient trap for self-interstitials.
Exploring this avenue, we demonstrate in this Letter, on the basis of ab initio calculations,
that
penta-, tetra-, and tri-vacancies  in the form of combinations of ring hexavacancies
with one, two, or three self-interstitials constitute very stable complexes,
with formation energies significantly lower  than ``part of hexagonal ring'' (PHR)
configurations.
In a sense, this family of defects is a
generalization of the ``fourfold coordinated point defect'' described in \cite{fourfold},
which is essentially a combination of a divacancy with two self-interstitials.

The calculations of the energies and relaxed geometries of the vacancy clusters
were performed using the Vienna Ab-initio Simulation Package (VASP), which
employs pseudopotential DFT with the projector augmented-wave method (PAW) \cite{Blochl, Kresse}.
We used a 216-atom supercell,
an energy cutoff of 22 Ry,
and the local-density approximation (LDA) for the exchange-correlation functional.
Results are reported for $\Gamma$-point sampling only of the Brillouin zone,
which we found is sufficient to ensure convergence of the relative energies
of the defects.
One of the standard experimental tools for the study of defects in semiconductors is
positron annihilation spectroscopy \cite{PAS}; we therefore also performed calculations of the
positron lifetimes for various vacancy clusters.
The positron wave-functions and annihilation rates were calculated using the
potential and electron density given by the DFT calculations; the effect of electron-positron correlations
was taken into account by introducing an additional correlation potential and annihilation
enhancement factor according to the interpolation formulas by Boro\'{n}ski and Nieminen
\cite{Boronski} with corrections for semiconductors
(see review by Puska and Nieminen \cite{pos_rev} for details).

\begin{figure*}
\scalebox{1.1}{\includegraphics{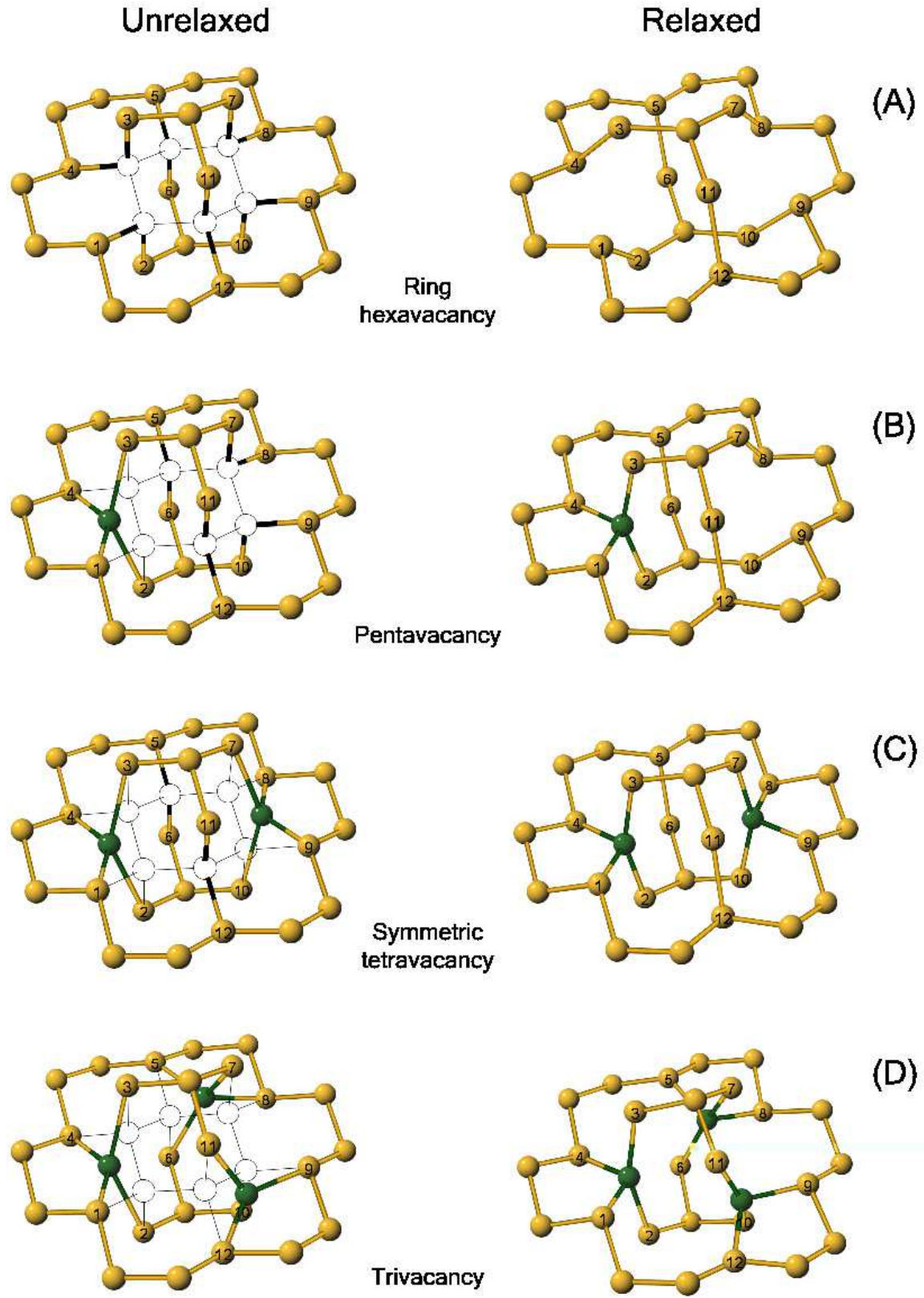}}
\caption{\label{fig:f1}
[Color online] Initial (left) and relaxed (right) geometries for the ring hexavacancy (A),
and for the pentavacancy (B), the
symmetric tetravacancy (C), and the trivacancy (D) in the fourfold configurations
(combinations of a hexavacancy with one, two or three self-interstitials, respectively).
The self-interstitials added to the hexavacancy are shown in dark green.
The open circles indicate the positions of the atoms removed from the lattice
to form the starting-point ring hexavacancy.
}
\end{figure*}

Figure \ref{fig:f1} presents the proposed fourfold configurations for
the penta-, tetra-, and tri-vacancy in silicon.
Figure \ref{fig:f1} (A) (left) shows the unrelaxed ring hexavacancy, with labels
1 to 12 indicating the twelve atoms each having initially one dangling bond.
In the process of relaxation, these twelve atoms form six new bonds with each other (right).
However, if one silicon atom is added to the defect, four of the
twelve atoms can form new bonds with it
while the others pair in the same way as in the case of
the simple hexavacancy [Fig.\ \ref{fig:f1} (B)].
If two atoms are added, they will satisfy eight of the twelve dangling bonds
while four atoms pair [Fig.\ \ref{fig:f1} (C)];
and if three atoms are added,
all twelve dangling bonds of the hexavacancy are satisfied [Fig.\ \ref{fig:f1} (D)].
Thus, these configurations for the penta-, the tetra-, and the
tri-vacancy have {\em no dangling bonds at all}.
As a result, they are expected to be more stable than the PHR
configurations where two dangling bonds remain at the ends of the vacancy chain after the lattice has relaxed.
Note that
for the tetravacancy, two different configurations are possible:
symmetric [Fig.\ \ref{fig:f1} (C)], where
the first atom is bonded to atoms  1 - 4 and the second
to atoms  7 - 10,
and non-symmetric, where the second atom is bonded to atoms  5 - 8.
For penta- and tri-vacancies, all possible configurations are equivalent.

\begin{table}
\caption{\label{tab:t1}
Calculated formation energies for various configurations of vacancy clusters
($N_v$ = number of vacancies), in eV per defect. In the PHR configurations,
atoms are removed sequentially from the hexagonal ring; the fourfold
configurations are shown in Fig.\ \ref{fig:f1}. For the latter, the symmetry
groups are also given.}
\begin{ruledtabular}
\begin{tabular}{llll}
 $N_v$ &  \multicolumn{2}{c}{Formation energy} & Group for  \\
       &  PHR confs. & Fourfold confs.         & fourfold confs. \\
\hline
1  &  3.51         &                 & \\
2  &  5.01         &                 & \\
3  &  6.80         &  6.20           & $D_3$ \\
4  &  8.26 (chain) &  7.26 (sym)     & $C_{2h}$ \\
   &  8.35 (PHR)   &  7.35 (non-sym) & $C_{1h}$ \\
5  &  9.07         &  8.42           & $C_2$ \\
6  &  9.41         &                 & $D_{3d}$ \\
\end{tabular}
\end{ruledtabular}
\end{table}

The expectation that the fourfold configurations described above are more stable than the usual
PHR configurations is verified by computing the formation energies.
In our calculations, each additional atom is initially placed
at the geometrical center of
its group of four future neighbours (see Fig.\ \ref{fig:f1}).
The system is then relaxed using the conjugate-gradient algorithm.
For comparison, we have also performed calculations for the PHR configurations and for the chain tetravacancy.

The calculated formation energies are presented in Table \ref{tab:t1}, where
we also give the symmetry groups of the fourfold defects. The formation
energies for the tri-, tetra-, and pentavacancies in the fourfold
configurations are indeed lower than those for the PHR configurations, by
0.6, 1.0, and 0.6 eV, respectively. It should be mentioned that, in contrast
to \cite{Hast} and \cite{Staab}, we find the chain tetravacancy to have lower
energy than the PHR configuration. However, both our calculations and
\cite{Hast, Staab} show a very small energy difference between the two
configurations so they can be considered equally stable.

As mentioned earlier, these results were obtained using only the
$\Gamma$-point to sample the Brillouin zone. In order to check for
convergence, we have also performed some calculations using a
$2\times2\times2$ Monkhors-Pack grid. We find that the formation energies
change by at most 0.5 eV, while the relative energies given above change by
no more than a few percent and are thus converged with respect to {\bf
k}-point sampling. Full convergence of the formation energies is numerically
intensive and would not alter our conclusions.

\begin{figure}
\scalebox{0.9}{\includegraphics{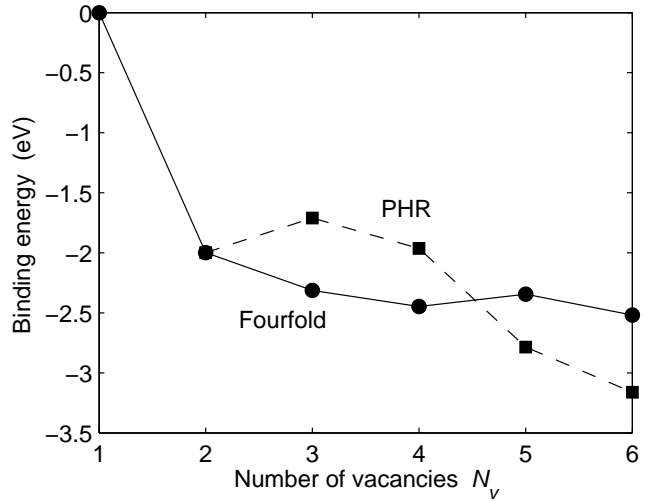}}
\caption{\label{fig:f3}
Binding energy for vacancy clusters as a function of size, $V_n \to V_{n-1}+V$.
The dashed line corresponds to PHR configurations
and the solid line to fourfold configurations (see Fig.\ \ref{fig:f1}).}
\end{figure}

Figure \ref{fig:f3} shows the calculated binding energy, i.e., the energy necessary
to remove one vacancy from a cluster, $V_n \to V_{n-1}+V$.
For the PHR configurations, our calculations are in good agreement with the results
of Staab {\em et al.} \cite{Staab}:
we also find the absolute value of the binding energy to be minimal for the trivacancy,
and to increase with the size of the cluster for $3 \leq N_v \leq 6$.
For the fourfold configurations, now, our calculations show the binding energy
to be approximately the same for all defect sizes.
For  $4 \leq N_v \leq 6$, this result can be explained by the structure of the defects:
additional atoms attach to the hexavacancy more or less independently,
and thus approximately the same energy is necessary to remove the first, second, or last atom.

In order to identify a possible formation mechanism for the fourfold vacancy clusters
(other than the capture of self-interstitials by an earlier-formed hexavacancy), we calculated,
using the nudged elastic band method \cite{neb},
the transition barrier between the PHR and the fourfold pentavacancy.
In this transition, an atom originally bonded to only
two neighbours (atoms 1 and 2 in Fig.\ \ref{fig:f1} (A), for example) moves to the interstitial position
to form two new bonds (with atoms 3 and 4 [Fig.\ \ref{fig:f1} (B)]). Our calculations show that the barrier for this transition is very low,
about 0.02-0.03 eV. Thus, the PHR pentavacancy should quickly move to the fourfold configuration.
In a similar way, a fourfold tetravacancy can easily be formed
from two parallel second-nearest-neighbour divacancies by moving two twofold atoms to the interstitial positions.
Likewise, a possible initial configuration for the formation of
the fourfold trivacancy is three second-nearest-neighbour vacancies in the hexagonal ring.

Table \ref{tab:t2} presents the calculated positron lifetimes for the different vacancy clusters.
The calculations were performed for both relaxed and unrelaxed geometries.
For the unrelaxed PHR configurations, our lifetimes are in perfect agreement
with the results of Staab {\em et al.} \cite{Staab}.
For the relaxed PHR states, we find close agreement
for larger clusters, whereas for
$N_v \leq 3$,
our calculations give values larger by 10-15 ps.
A possible explanation for the (small) differences is some variations of the relaxed geometries
due to the use of different methods of calculation.

It is generally believed that the lifetimes calculated for unrelaxed
geometries correspond more closely to experiment because positron-induced
outward relaxation compensates for the usual inward relaxation around vacancy
clusters \cite{Staab, Saito}. However, in the case of the ring hexavacancy,
such calculations overestimate the lifetime by about 20 ps \cite{Saito}.
Moreover, it is not clear what ``unrelaxed geometries'' means for the
fourfold configurations. In our calculations, we simply take this as the
initial configuration where the interstitial atoms are placed at the
geometrical centers of their groups of neighbours. One can see from Fig.\
\ref{fig:f1} that the additional atoms really move towards the defect center
from their initial positions in the process of relaxation, which makes our
choice reasonable. Obviously, calculations performed for the geometries
relaxed with respect to both electronic and positronic forces are necessary
to get reliable values of the lifetimes for the fourfold configurations.
Nevertheless, the numbers shown in Table \ref{tab:t2} provide reasonable
estimates.

\begin{table}
\caption{\label{tab:t2}
Calculated positron lifetime (in ps) for the unrelaxed and relaxed (in brackets) geometries of
the same vacancy clusters as in Table \ref{tab:t1}.
}
\begin{ruledtabular}
\begin{tabular}{lll}
 $N_v$ &  PHR confs. & Fourfold confs.\\
\hline
1  &  252 (226)         &              \\
2  &  296 (255)         &              \\
3  &  329 (290)         &  321 (258)         \\
4  &  343 (291) (chain) &  342 (292) (sym)   \\
   &  340 (294) (PHR)   &  347 (298) (non-sym)\\
5  &  354 (301)         &  363 (312)         \\
6  &  376 (316)         &              \\
\end{tabular}
\end{ruledtabular}
\end{table}

We have performed electronic structure calculations and found that, like the
simple ring-hexavacancy \cite{Hast, Estr}, fourfold vacancy clusters have no
energy levels in the band gap. As a result, they should be optically
inactive, making their direct observation difficult. That being said, there
is experimental evidence, mostly from positron annihilation spectroscopy,
that the fourfold configurations are likely states of these defects.
In particular, Motoko-Kwete {\em et al.} \cite{Motoko}
reported a positron lifetime value of 350 ps, consistent with both fourfold
and PHR configurations. However, they observe the defects to be more stable
at high temperature than the usual tetravacancies. This result has been
explained by the presence of impurites in the material. Our calculations suggest that
the defects actually are the fourfold configurations reported in this Letter.
Also, the formation of fourfold trivacancies provides a natural explanation
to the experimental results of Poirier {\em et al.} \cite{Poirier}; these authors
have observed that, in the process of divacancy annealing at $T=250$ C, the
infrared absorption, which is associated with divacancies, decreases with
time, while positron lifetime and trapping rate remain unchanged. According
to Table \ref{tab:t2}, the difference between positron lifetimes for fourfold
trivacancies and divacancies is rather small, certainly within the
uncertainty arising from the computational method (see above). Since
fourfold trivacancies are invisible to infrared spectroscopy, the
``coalescence'' of divancies into fourfold trivacancies resolves the apparent
contradiction reported in \cite{Poirier}.

In summary, we propose new fourfold configurations for tri-, tetra-, and penta-vacancies in silicon.
Our DFT calculations show that they have formation energies lower by 0.6, 1.0, and
0.6 eV, respectively, than the PHR
configurations, generally believed to be the stable states of these defects.
We have identified a possible formation mechanism for the fourfold vacancy clusters
and performed preliminary calculations for positron lifetimes associated with them.

We are grateful to R\'emi Poirier for useful discussions.
This work was supported by grants from the
Natural Sciences and Engineering Research Council (NSERC) of Canada and
the ``Fonds Qu{\'e}b{\'e}cois de la
recherche sur la nature et les technologies'' (FQRNT) of the Province of
Qu{\'e}bec. We are indebted to the ``R\'eseau qu\'eb\'ecois de calcul de
haute performance'' (RQCHP) for generous allocations of computer
resources.

\bibliographystyle{apsrev}

\end{document}